\documentclass
[prd,twocolumn,floatfix,amsmath,nofootinbib,amssymb,floatfix]{revtex4}
\usepackage{graphicx,color,dcolumn,booktabs,bm}
\usepackage{longtable,lscape}
\usepackage{txfonts}
\usepackage{overpic}
\usepackage{amssymb}
\usepackage{array}
\usepackage{indentfirst}
\usepackage{feynmf}   
\usepackage{slashed}  
\usepackage{cases}
\usepackage{color}
\usepackage{multirow}
\usepackage{epstopdf}
\usepackage{tabularx}
\usepackage[compat=1.1.0]{tikz-feynman}
\usepackage[colorlinks,
            citecolor=blue,
            anchorcolor=red,
            menucolor=red,
            linkcolor=red,
            filecolor=red,
            runcolor=red,
            urlcolor=blue,
            frenchlinks=red]{hyperref}

\begin{document}
\title{Productions of bottom and bottom-strange mesons in pion and kaon induced reactions}
\author{Jing Liu$^1$}\email{jingliu@seu.edu.cn}
\author{Quan-Yun Guo$^1$}\email{qyguo@seu.edu.cn}
\author{Qi Wu$^2$}\email{wuqi@htu.edu.cn}
\author{Jun He$^{3,4}$}\email{junhe@njnu.edu.cn}
\author{Dian-Yong Chen $^{1,4}$ \footnote{Corresponding author}} \email{chendy@seu.edu.cn}
\affiliation{
$^1$ School of Physics, Southeast University,  Nanjing 210094, China\\
$^2$ Institute of Particle and Nuclear Physics, Henan Normal University, Xinxiang 453007, China\\
$3$ Department of Physics and Institute of Theoretical Physics,
Nanjing Normal University, Nanjing 210097, China\\
$^4$  Lanzhou Center for Theoretical Physics, Lanzhou University, Lanzhou 730000, China
}

\begin{abstract}
In the present work, we propose to explore the productions of the bottom and bottom-strange mesons in the high-energy pion and kaon-induced reactions on a proton target. The cross sections are evaluated with an effective Lagrangian constructed by the heavy-quark limit and chiral symmetry. Our estimations show that at $P_\pi=80$ GeV, the cross sections for $B(5279)$, $B^\ast (5325)$, $B_0^\ast (5738)$, $B_1^\prime (5757)$, $B_1(5721)$ and $B_2^\ast (5747)$ production processes are estimated to be $3.19 \sim 86.26$, $1.86\sim 51.29$, $0.87 \sim 24.25$, $0.84 \sim 23.14$, $162.35 \sim 4477.66$, and $57.16 \sim 1604.43$ nb, respectively, where uncertainties arise from the model parameter. In addition, the cross sections for the corresponding bottom-strange mesons production processes are very similar. Moreover, our estimations indicate that the ratios of these cross sections are almost independent on the model parameters. In particular, the cross-section ratios related to the states in the same doublets are of order one, which is consistent with the expectation of heavy-quark limit. The cross sections related to the states in the $T$ doublets are about two orders larger than those related to the states in the $S$ doublets.

\end{abstract}


\maketitle

\section{Introduction}
\label{sec:introduction}
The bottom quark was first observed in 1977, and a di-muon resonance at about 9.5 GeV was observed in proton-nuclues collision at the Fermi Lab~\cite{E288:1977xhf}, as well as in $e^+ e^-$ annihilation at DESY~\cite{Pluto:1978tuc}. The first two $b$ flaved mesons, $B^+$ and $B^0$, were reported by the CLEO Collaboration in the $\Upsilon(4S)$ strong decay in 1983~\cite{CLEO:1983mma}, while the first $b$ flavor meosn with strangeness, $B_s^0$, was discovered by the ALEPH~\cite{ALEPH:1993mpb} and CDF~\cite{{CDF:1993pzh}} Collaborations in 1993. As important members of the heavy-light meson system, the bottom and bottom-strange meson families are particular interesting since the bottom quark is heavy enough compared to the charmed quark. Thus, the bottom and bottom-strange mesons provide us with an excellent platform for testing the heavy-quark effective theory.

Even 30 years after the discovery of the first bottom meson, the members of the bottom meson family are still scarce. Only seven bottom mesons are listed in the Review of Particle Physics~\cite{ParticleDataGroup:2022pth}, which are $B(5279)$, $B^{\ast}(5325)$, $B_1(5721)$, $B_J(5732)$, $B_2(5747)$, $B_J(5840)$, and $B_J(5970)$, respectively. The first two bottom mesons, $B(5279)$ and $B^{\ast}(5325)$, have been well identified as $1S$ states with $J=0$ and $J=1$, respectively. In 1994, the OPAL Collaboration observed evidence for the kinematic and charge correlations of $B$ mesons with charged pions using the data collected by the OPAL detector at LEP, and an excess in the $B^+ \pi^-$ invariant mass distribution in the range $5.60-5.85$ GeV was reported~\cite{OPAL:1994hqv}. This structure, named $B_J(5732)$, has been confirmed by some subsequent experimental measurements~\cite{DELPHI:1994fnu,ALEPH:1995ikc, ALEPH:1998unp,L3:1999pdo,CDF:1999zui}, and it has been interpreted as stemming from several narrow and broad bottom mesons. In Ref.~\cite{D0:2007vzd}, the $P$-wave bottom states, $B_1(5721)$ and $B_2(5747)$, were observed directly for the first time in the decays to $B^{+(\ast)} \pi$ by the D0 Collaboration in 2007. These two states had also been observed by the CDF Collaboration in 2008~ \cite{CDF:2008qzb}. Using the complete CDF Run II data sample, the CDF Collaboration further confirmed the existence of $B_1(5721)$ and $B_2(5747)$ in 2013, and additionally reported the evidence for a new resonance, named $B(5970)$~\cite{CDF:2013www}. In 2014, the LHCb Collaboration analyzed the invariant mass distributions of $B^+ \pi^-$ and $B^0 \pi^+$, and the resonance parameters of $B_1(5721)$ and $B_2(5747)$ had been precisely measured. In addition, clear enhancements were observed over background in the mass range $5800\sim 6000$ MeV, and these structures were identified as $B_J(5840)$ and $B_J(5900)$, respectively~\cite{LHCb:2015aaf}.

Similarly to the case of the bottom meson, the members of the bottom-strange meson family are not abundant, either. There are also seven bottom-strange mesons in the Review of Particle Physics~~\cite{ParticleDataGroup:2022pth}, which are $B_s (5367)^0$, $B_s^{\ast} (5415)^0$, $B_{s1}(5830)^0$, $B_{s2}^{\ast}(5840)^0$, $B_{sJ}^\ast (5850)^0$, $B_{sJ}(6063)^0$, and $B_{sJ}(6114)^0$. The ground $S-$wave states, $B_s (5367)^0$ and $B_s^{\ast} (5415)^0$, have been well established. Similarly to the case of $B_J(5732)$, the OPAL Collaboration also reported the evidence of the structure, named as $B_{sJ}^\ast (5850)^0$, in the $B^+K^-$ invariant mass distribution in the range $5.80-6.00$ GeV in Ref.~\cite{OPAL:1994hqv}. Using the $p\bar{p}$ collision at $\sqrt{s}=1.96$ TeV collected with the CDF II detector at the Fermilab Tevatron, the CDF Collaboration reported the first observation of the narrow $j_q=3/2$ states of the orbitally excited bottom-strange meson, which were $B_{s1}(5830)^0$ in the $B^{\ast+} K^-$ channel and $B_{s2}^{\ast}(5840)^0$ in the $B^+K-$ channel, respectively \cite{CDF:2007avt}. Almost at the same time, the D0 Collaboration also reported the direct observation of the excited $L=1$ state $B_{s2}^{\ast}(5840)^0$ in the invariant mass distribution of $B^+K^-$  with a statistical significance of more than $4.8\sigma$~\cite{D0:2007die}. In addition, the presence of the $B_{s1}$ signal was also tested and the significance of $B_{s1}(5830)^0$ was reported to be less than $3 \sigma$~\cite{D0:2007die}. In Ref.~\cite{LHCb:2012iuq}, the LHCb collaboration reported the first observation of the decay process $B_{s2}^{\ast}(5840)^0\to B^{\ast +} K^-$. Later, the CMS Collaboration measured the $B_{s2}^{\ast}(5840)^0$ and $B_{s1}(5830)^0$ meson in the $B^{(\ast) +} K^-$ and $B^{(\ast) 0} K_s^0$, and the branching fraction of the neutral channel relative to the charged channel for $B_{s2}^{\ast}(5840)^0$ was measured~\cite{CMS:2018wcx}. The last two previously discovered bottom-strange mesons, $B_{sJ}(6063)^0$, and $B_{sJ}(6114)^0$, were observed in the $B^{\pm} K^{\mp}$ invariant mass distributions by the LHCb Collaboration in 2020 ~\cite{LHCb:2020pet}.

After the observations of the bottom and bottom-strange mesons, some theoretical investigations have been performed from various aspects. In some kinds of quark model, for instance, the non-relativistic quark model~\cite{Isgur:1998kr, Asghar:2018tha}, relativistic quark model~\cite{Ebert:1997nk,Liu:2013maa,Liu:2015lka, Wang:2016itc}, MIT bag model~\cite{Orsland:1998de}, the semi-relativistic quark potential model~\cite{Matsuki:2006zoi}, chiral quark model~\cite{Matsuki:2006zoi, DiPierro:2001dwf, Xiao:2014ura}, and the modified relativistic quark model~\cite{Sun:2014wea}, the mass spectra of bottom or bottom-strange mesons as well as the decay properties had been investigated. In addition to the quark model, the effective Lagrangian approach had also been employed to investigate the properties of the bottom and bottom-strange mesons. In Ref.~\cite{Falk:1995th}, the decays of excited bottom and bottom-strange mesons were calculated by including the leading corrections to the heavy-quark limit. The masses and widths of $P$ wave bottom and bottom-strange mesons were estimated using the heavy-quark symmetry supplemented by insights gleaned from potential models~\cite{Eichten:1993ub}. However, in Ref.~\cite{Lewis:2000sv}, the mass spectra of $S$ and $P$-wave bottom and bottom-strange mesons were computed in the quenched approximation by using NRQCD up to third order in the inverse heavy-quark mass expansion. Adopting the effective Lagrangian approach based on the heavy quark and chiral symmetry, the branching fractions of some bottom and bottom-strange meson were evaluated in Refs.~\cite{Colangelo:2006kx, Colangelo:2012xi,Wang:2014cta}. Additionally, heavy-light meson spectra were investigated from lattice QCD using static heavy quarks~\cite{Green:2003zza}. The spectra and the $M_1$ transition widths of the bottom and bottom-strange mesons are calculated within the framework of the Blankenbecler-Sugar equation in Ref.~\cite{Lahde:1999ih}.

Compared with charmed or charmed strange mesons, the observations of the bottom and bottom-strange mesons are not abundant. At present, charmed and charmed strange mesons could be produced in the $e^+ e^- $ annihilation~~\cite{BaBar:2003cdx,SELEX:2004drx,CLEO:2003ggt,Belle:2003kup, Belle:2003guh, BaBar:2004yux,BaBar:2006eep}, $B$ meson decays~\cite{BaBar:2003oey,Datta:2003re}, $pp/p\bar{p}$ collisions~\cite{LHCb:2013jjb}, $\gamma A$ collision~\cite{FOCUS:2003gru}, $e^\pm p$ collisions~\cite{ZEUS:2012gyr},  $\pi A$ collision~\cite{ACCMOR:1990xso}, and even neutrino collision~\cite{BigBubbleChamberNeutrino:1995amd} processes. But for the bottom and bottom-strange meson productions, the production processes are not as abundant as the charmed and charmed strange meson due to their high masses, which makes the experimental research for bottom or bottom-strange mesons limited. In the current study, we proposed to explore the production of bottom and bottom-strange mesons in the  pion and kaon-induced reactions with the high-energy pion/kaon beam, which may be potentially accessible at facilities such as J-PARC~\cite{Nagae:2008zz}, OKA@U-70~\cite{Obraztsov:2016lhp}, COMPASS~\cite{Nerling:2012er},  SPS@CERN~\cite{Wise:1992hn} and AMBER@CERN~\cite{Quintans:2022utc}. The cross sections for the productions of $S$ and $P-$ wave bottom and bottom-strange mesons in the $\pi p$ and $Kp$ collisions are estimated by using an effective Lagrangian approach, which is constructed by the heavy-quark limit and chiral symmetry. The results in the present work could offer some useful information for further experimental measurements, and on the other hand, the experimental measurements could in turn provide crucial tests to the present estimations and the heavy-quark effective theory as well.

This work is organized as follows. After the introduction, we introduce the theoretical framework used in the present estimations. In Section ~\ref{Sec:Num}, the cross sections, differential cross sections for the considered processes and the relevant discussions are shown, and the last section is devoted to a short summary.

\section{Pion (Kaon) induced production on a proton target}
\label{sec:APPROACH}

\renewcommand\arraystretch{1.25}
\begin{table}[t]
 \centering
 \caption{The $S$- and $P$- wave bottom and bottom-strange mesons. \label{Tab:CP}}
 \begin{tabular}{p{1.1cm}<\centering p{1.2cm}<\centering p{1.0cm}<\centering p{2.1cm}<\centering  p{2.1cm}<\centering}
 \toprule[1 pt]
Doublet & $J^P$  &  State &   $b\bar{q}$ & $b\bar{s}$ \\
 \midrule[1 pt]
$H$& $0^-$ &    $P$     &  $B(5279)$  \cite{ParticleDataGroup:2022pth}       &$B_{s}(5366)$  \cite{ParticleDataGroup:2022pth} \\
       & $1^-$ &    $P^*$     &  $B^*(5325)$  \cite{ParticleDataGroup:2022pth}         &$B^*_{s}(5415)$ \cite{ParticleDataGroup:2022pth} \\
$S$&  $0^+$&   $P^*_0$    &$B_0^*(5738)$\cite{Ebert:2001zm}  &$B^*_{s0}(5841)$\cite{Ebert:2001zm} \\
      & $1^+$&  $P^{\prime}_1$  &$B_1^{\prime}(5757)$ \cite{Ebert:2001zm}      &$B^{\prime}_{s1}(5859)$\cite{Ebert:2001zm} \\
$T$&  $1^+$ &   $P_1$       &$B_1(5721)$ \cite{ParticleDataGroup:2022pth}      &$B_{s1}(5830)$ \cite{ParticleDataGroup:2022pth}  \\
      &$2^+$ &  $P^*_{2}$    &$B_2^*(5747)$ \cite{ParticleDataGroup:2022pth}     &$B^*_{s2}(5840)$  \cite{ParticleDataGroup:2022pth} \\
  \bottomrule[1 pt]
 \end{tabular}
 \end{table}
In the present work, we discuss the $S$- and $P$-wave bottom and bottom-strange mesons productions in the pion and kaon induced productions on proton target. In Table \ref{Tab:CP}, we collect the relevant $S-$ and $P$-wave bottom/bottom-strange mesons, where $B_{0}^\ast(5738)$, $B_1^\prime(5757)$, $B_{s0}^\ast(5841)$,  and $B_{s1}^\prime(5859)$ have not been observed yet, and their masses used in the present estimations are taken from the quark model predictions in Ref.~\cite{Ebert:2001zm}. The present estimations for the productions of these unobserved states may provide some useful information for searching them experimentally.


\subsection{Effective Lagrangians}
For the heavy-light system, the total angular momentum $\vec{J}$ can be decoupled as $\vec{J}=\vec{S}_{Q}+\vec{S}_\ell$, where $\vec{S}_{Q} $ is the spin of the heavy quark, while $\vec{S}_\ell=\vec{s}_{\bar{q}} +\vec{\ell}$ is the light degree of freedom with $\vec{s}_{\bar{q}} $ and $\vec{\ell}$ to be spin of the light antiquark and the orbital angular momentum, respectively. In the heavy quark limit, the heavy quarks are infinitely heavy, which lead to the spin of the heavy quark $\vec{S}_{Q}$ and the total angular momentum of the light degrees of freedom $\vec{s}_\ell$ are separately conserved by the strong interactions.  For the $S$-wave states, the orbital angular momentum $\ell=0$ and their spin parities could be $J^P=0^+$ and $1^+$, which refer to $P$ and $P^{*}$ in Table \ref{Tab:CP}, respectively. In the heavy quark effective theory, these two states are degenerated and could be expressed by the superfield $H_{a}^{(Q)}$ in the matrix form, which is,
\begin{eqnarray}
H_{a}^{(Q)}&=&\frac{1+v\!\!\!\slash}{2} \left[P_{a}^{*(Q)\mu}\gamma_{\mu}-P^{(Q)}_{a}\gamma_{5}\right],
\end{eqnarray}
where $Q$ and $a$ represent the heavy quark and the light flavor index $(u,\ d,\ s)$, respectively.

In the similar manner, the $P$-wave heavy-light mesons can be categorized into two doublets, which are the $S$ doublet with $s_\ell=1/2$ and $T$ doublet with $s_\ell =3/2$, respectively. In the $S$ doublet, the $J^P$ quantum numbers for the two states are $0^+$ and $1^+$, corresponding to $P_0^\ast$ and $P_1^\prime$ listed in Table \ref{Tab:CP}. Meanwhile, the $T$ doublet consists of the states $P_1$ and $P_2^\ast$ with $J^P=1^+$ and $2^+$, respectively. The matrix expressions of the superfields $S$ and $T$ are defined as,
\begin{eqnarray}
S^{(Q)}_{a}&=&\frac{1+v\!\!\!\slash}{2} \left[P^{'(Q)\mu}_{1a}\gamma_{\mu}\gamma_{5}
        -P^{*(Q)}_{0a} \right] ,\nonumber\\
T^{(Q)\mu}_{a}&=&\frac{1+v\!\!\!\slash}{2} \left[P^{*(Q)\mu\nu}_{2a}\gamma_{\nu}-
          \sqrt{\frac{3}{2}}P^{(Q)}_{1a\nu}\gamma_{5} \left (g^{\mu\nu}
           -\frac{1}{3}\gamma^{\nu}(\gamma^{\mu}-v^{\mu}) \right) \right ]. \nonumber\\
\end{eqnarray}

Considering the chiral symmetry, the coupling between chiral particles and heavy-light mesons can be constructed as,
 \begin{eqnarray}
 \mathcal{L}_{\rm int}&=&ig\langle H  A\!\!\!\slash \gamma_{5} \bar{H}\rangle +ih \left\langle S {A}\!\!\!\slash\gamma_{5}\bar{H}\right\rangle\nonumber\\&&
             +i\frac{h^\prime}{\Lambda_{\chi}} \left\langle T^{\mu}\left (D_{\mu}\ A\!\!\!\slash +D\!\!\!\!\slash \  A_{\mu}\right)\gamma_{5}\bar{H}\right \rangle , \label{Eq:LagA}
  \end{eqnarray}
  where $D_{\mu}=\partial_{\mu}+V_{\mu}$ is covariant derivative. $V_{\mu}$ and $A_{\mu}$ are the vector and axial currents, which are defined as,
\begin{eqnarray}
V_\mu &=& \frac{1}{2} \left[\xi^\dagger \partial_\mu \xi +\xi \partial_\mu \xi^\dagger \right] ,\nonumber\\
A_\mu &=& \frac{1}{2} \left[\xi^\dagger \partial_\mu \xi -\xi \partial_\mu \xi^\dagger \right] ,\nonumber
\end{eqnarray}
with $\xi$ = $\exp{\left(iM/f_{\pi}\right)}$ to be the nonlinear representation of Goldstone field, and $M$ is the matrix of pseudoscalar meson octet, which is,
\begin{eqnarray}
 {M}&=&\left(
               \begin{array}{ccc}
                 \sqrt{\frac{1}{2}}\pi^{0}+ \sqrt{\frac{1}{6}}\eta & \pi^{+} & K^{+} \\
                 \pi^{-} &  -\sqrt{\frac{1}{2}}\pi^{0}+ \sqrt{\frac{1}{6}}\eta & K^{0} \\
                 K^{-} & \bar{K}^{0} & -\sqrt{\frac{2}{3}}\eta \\
               \end{array}
             \right).
 \end{eqnarray}

By expanding Eq.~(\ref{Eq:LagA}), we obtain the Lagrangians involved in the present estimations, which are,
\begin{eqnarray}
\mathcal{L}_{P^{*}PM}&=&g_{P^{*}PM}\left(P_{\beta}^{*}\partial^{\beta}MP^{\dag}
            +P\partial^{\beta}MP_{\beta}^{*\dag}\right)+h.c. ,\nonumber\\
\mathcal{L}_{P^{*}P^{*}M}&=&ig_{P^{*}P^{*}M}\varepsilon^{\beta\lambda\mu\sigma}
                            P^{*}_{\mu}
                           {\stackrel{\leftrightarrow}{\partial}}_{\lambda}
                           P^{*\dag}_{\beta}
                           \partial_{\sigma}M+h.c. , \nonumber\\
 \mathcal{L}_{P_{0}^{*}PM}&=&ig_{P_{0}^{*}PM}\left(P^{*}_{0}
           {\stackrel{\leftrightarrow}{\partial}}_{\mu}P^{\dag}\right)\partial^{\mu}M+h.c. ,
       \nonumber\\
\mathcal{L}_{P_{1}^{'}P^{*}M}&=&ig_{P_{1}^{'}P^{*}M} \left(P^{'\mu}_1
            {\stackrel{\leftrightarrow}{\partial}}_{\nu} P_{\mu}^{*\dagger}\right)\partial^{\nu}M+h.c. ,\nonumber\\
\mathcal{L}_{P_{1}P^{*}M}&=&g_{P_{1}P^{*}M}\left[3P^{\mu}_{1}\left(\partial_{\mu}
            \partial_{\nu}M\right)P^{*\nu\dagger}-P^{\mu}_{1}
            (\partial^{\nu}\partial_{\nu}M)P^{*\dag}_{\mu}\right]+h.c. ,\nonumber\\
\mathcal{L}_{P^{*}_{2}PM}&=&  g_{P_2^\ast P M}P^{*\mu\nu}_{2}\left( \partial_{\mu}\partial_{\nu} M\right)P^{\dagger}+h.c. ,    \nonumber\\
\mathcal{L}_{P^{*}_{2}P^{*}M}&=&ig_{P^{*}_{2}P^{*}M}
            \varepsilon^{\alpha\lambda\eta\sigma}
            \left(P^{*\mu}_{2\eta}{\stackrel{\leftrightarrow}{\partial}}_\sigma
            P^{*\dag}_{\alpha}\right)\left(\partial_\mu\partial_{\lambda} M\right)+h.c..
\end{eqnarray}

 In the conditions of heavy quark limit and chiral symmetry, the coupling constants in the above Lagrangians satisfy,
\begin{eqnarray}
 g_{P^{*}PM}&=&-\frac{2g}{f_{\pi}}\sqrt{m_{P}m_{P^{*}}} \ ,\nonumber\\
 g_{P^{*}P^{*}M}&=&\frac{g}{f_{\pi}}\  ,\nonumber\\
 g_{P_{0}^{*}PM}&=&  g_{P_{1}^{'}P^{*}M} =-\frac{h}{f_{\pi}}\  ,\nonumber\\
 g_{P_{1}P^{*}M}&=&-2\sqrt{\frac{2}{3}}\frac{h'}{\Lambda_{\chi}f_{\pi}}
            \sqrt{m_{P_{1}} m_{P^{*}}}\   ,  \nonumber\\
 g_{P^{*}_{2}P^{*}M}&= &  -\frac{2h^{\prime}}{\Lambda_{\chi}f_{\pi}} \ , \nonumber\\
 g_{P^{*}_{2}PM}&=& \frac{4h^{\prime}}{\Lambda_{\chi}f_{\pi}}\sqrt{m_{P_{2}^{*}}m_{P}}\ ,
\end{eqnarray}
with $ f_{\pi}=132$ MeV to be the decay constant of pion, $h=0.56\pm0.04$ and $h^{\prime}=0.43\pm0.01$. $\Lambda_{\chi}= 1$ GeV is the chiral symmetry-breaking scale.

Besides the above effective Lagrangian related to the heavy-light mesons and the chiral particles, the effective coupling between the baryons and mesons are also needed, and the involved effective Lagrangians read,
  \begin{eqnarray}
     \mathcal{L}_{\Lambda^{0}_{b}pB}&=&ig_{\Lambda^{0}_{b}pB}\bar{\Lambda}^{0}_{b}
            \gamma_{5}pB+h.c. \nonumber\\
     \mathcal{L}_{\Lambda^{0}_{b}pB^{\ast}}&=&g_{\Lambda^{0}_{b}pB^{\ast}}
            \bar{\Lambda}^{0}_{b} \gamma^{\mu}pB_{\mu}^{*}+h.c.,
  \end{eqnarray}
with  $g_{\Lambda^{0}_{b}pB}=13.1$ and
$g_{\Lambda^{0}_{b}pB^{\ast}}=4.3$~\cite{He:2016pfa,Dong:2014ksa}.

  \begin{figure*}[htb]
\begin{tabular}{cccc}
  \centering
  \includegraphics[width=4.00cm]{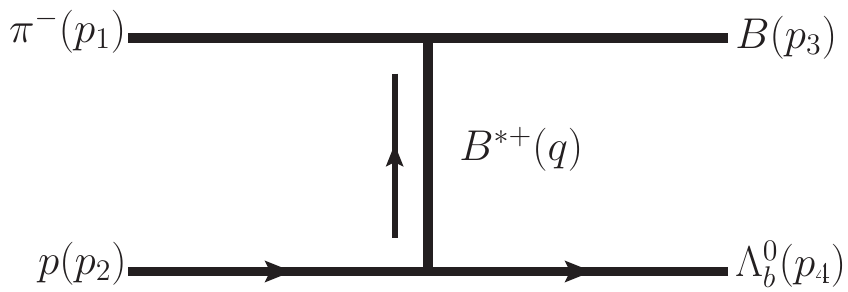}&
  \includegraphics[width=4.00cm]{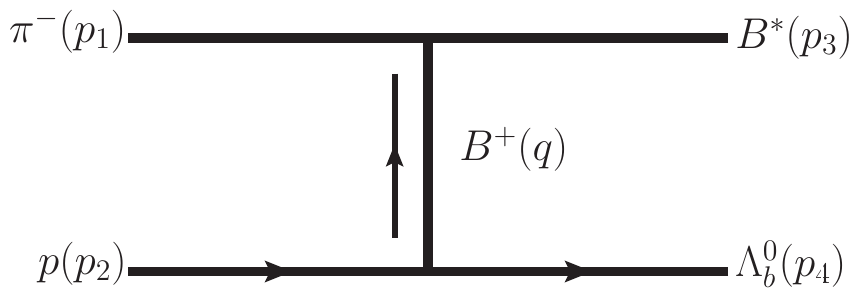}&
  \includegraphics[width=4.00cm]{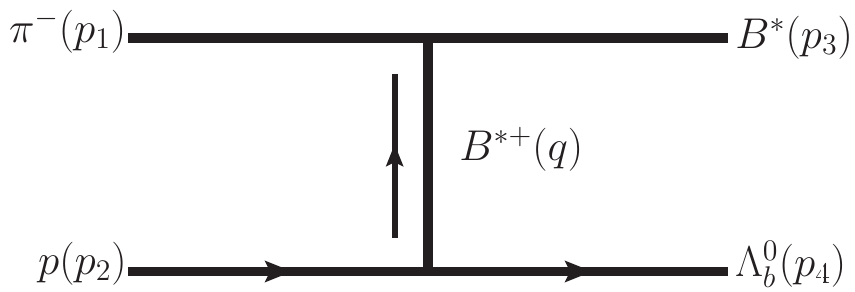}&
 \includegraphics[width=4.00cm]{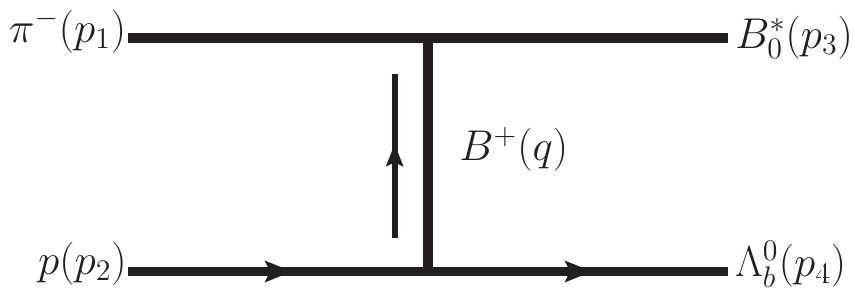}
 \\
 $\rm(a)$ & $\rm(b)$&
 $(c)$ & $(d)$\\
 \\
 \includegraphics[width=4.00cm]{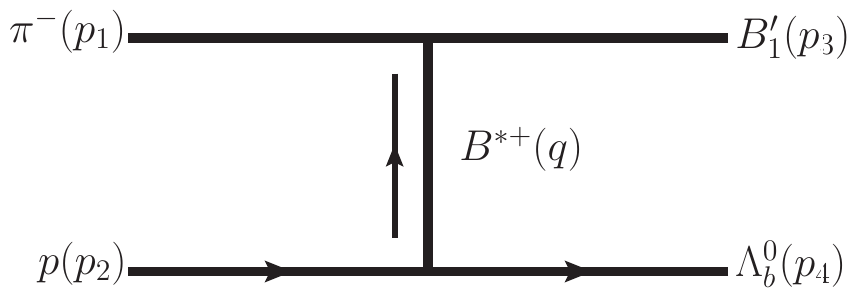}&
 \includegraphics[width=4.00cm]{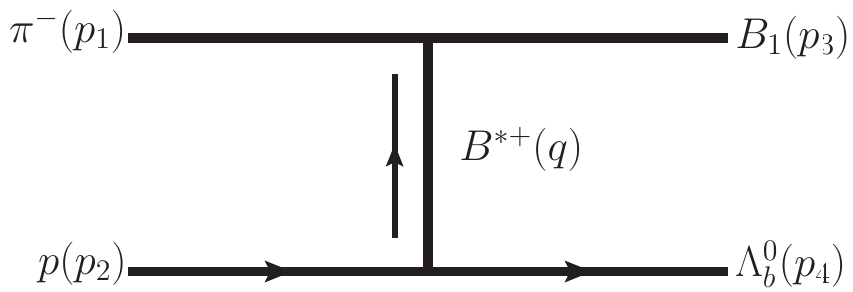}&
 \includegraphics[width=4.00cm]{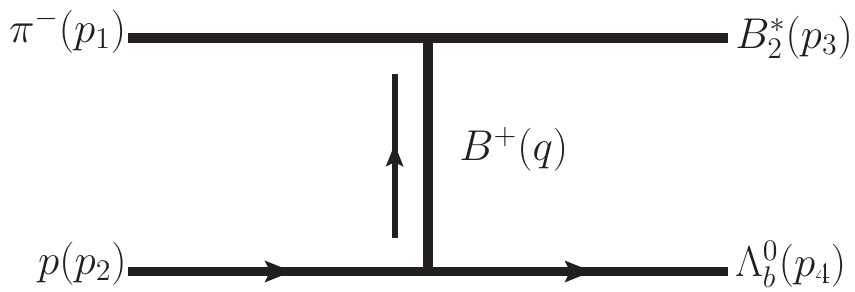}&
 \includegraphics[width=4.00cm]{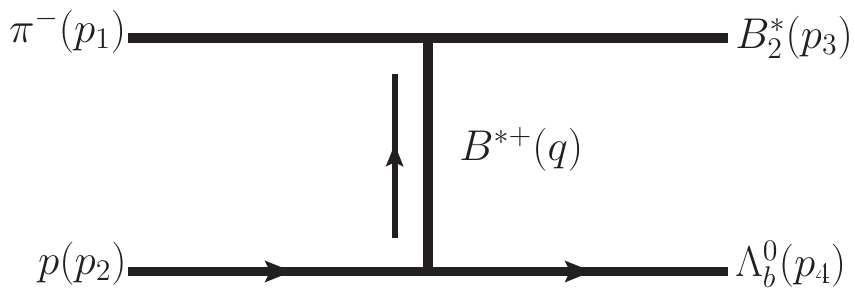}\\
 $\rm(e)$ & $\rm(f)$&
 $\rm(g)$ & $\rm(h)$\\
 \\
 \end{tabular}
 \caption{ Feynman diagrams for
 $\pi^- p \to B^{0} \Lambda^{0}_b$ (digram (a)),
 $\pi^- p \to B^{*0}  \Lambda^{0}_b$ (diagrams (b) and (c)),
 $\pi^- p \to B^{*}_{0}  \Lambda^{0}_b$ (diagram (d)),
 $\pi^- p \to B^{\prime}_{1}  \Lambda^{0}_b$ (diagram (e)),
 $\pi^- p \to B_{1}  \Lambda^{0}_b$ (diagram (f)),
 $\pi^- p \to B_{2}^{*0} \Lambda^{0}_b$(diagrams (g) and (h)).
  The diagrams associated with the $S$- and $P$- wave bottom-strange meson production processes could be obtained by replacing $\pi^-$ with $K^-$.  \label{Fig:Feyn}}
\end{figure*}

  \subsection{Amplitudes For the Scattering Processes}

In the present work, we focus mainly on the production of the $S$ and $P$ wave bottom (strange) mesons in the pion (kaon) induce reactions. The production processes could occur by exchanging a $B$ or a $B^\ast$ meson as indicated in Fig.~\ref{Fig:Feyn}. With the above effective Lagrangians, we could obtain the amplitudes corresponding to the diagrams in Fig.~\ref{Fig:Feyn}, which are,

\begin{eqnarray}
\mathcal{M}_{a}&=&\Big[g_{\Lambda^{0}_{b}B^{*}p}\bar{u}(p_{4})\gamma^{\mu}u(p_{2})\Big]
              \Big[g_{B^{*}B^{0}\pi}(-ip_{1}^{\nu}) \Big]
              \nonumber\\
              &&\times
              \frac{-g_{\mu\nu}+q_{\mu} q_{\nu}/m_{B^{*}}^{2}}{q^{2}-m_{B^{*}}^{2}}    F^{2}\left(q^{2},m^2_{B^{*}}\right) ,\nonumber\\
\mathcal{M}_{b}&=&\Big[g_{\Lambda^{0}_{b}Bp}\bar{u}(p_{4})\gamma_{5}u(p_{2})\Big]
             \Big[g_{B^{*0}B\pi}ip_{1\beta}\epsilon_{\beta}(p_{3})\Big]
              \nonumber\\
             &&\times
              \frac{1}{q^{2}-m_{B}^{2}}
              F^{2}\left(q^{2},m_{B}^2\right) ,\nonumber\\
              \mathcal{M}_{c}&=&\Big[g_{\Lambda_{b}^{0}B^{*}p}\bar{u}(p_{4})\gamma^{\mu}u(p_{2})\Big]
              \nonumber\\
              &&\times
              \Big[ig_{B^{*}B^{*}\pi}\epsilon^{\beta\lambda\xi\sigma}\left(-ip_{3\lambda}-iq_{\lambda}\right)
              \epsilon_{\beta}(p_{3})(-ip_{1\sigma})\Big]
              \nonumber\\
              &&\times
              \frac{-g_{\mu\xi}+q_{\mu}q_{\xi}/m^{2}_{B^{*}}}{q^{2}-m^{2}_{B^{*}}}
              F^{2}\left(q^{2},m^2_{B^{\ast}}\right) ,\nonumber
\end{eqnarray}
\begin{eqnarray}
 \mathcal{M}_{d}&=&\Big[g_{\Lambda^{0}_{b}pB}\bar{u}(p_{4})\gamma_{5}
               u(p_{2})\Big]  \Big[-ig_{B^{*}_{0}B\pi}(iq^{\mu}+ip_{3}^{\mu})(ip_{1\mu})\Big]\nonumber\\
           &&\times
           \frac{1}{q^{2}-m_{B}^{2}}F^{2}\left(q^{2},m_{B}^2\right) ,\nonumber\\
\mathcal{M}_{e}&=&\Big[g_{\Lambda^{0}_{b}pB^{*}}\bar{u}(p_{4})
         \gamma_{\beta}
         u(p_{2})\Big]\nonumber\\ &&\times\Big[ g_{B_{1}B^{*}\pi}(iq_{\nu} ip_{1}^{\nu}
           +ip_{3\nu}ip_{_{1}}^{\nu})\epsilon_{\mu}(p_{3})\Big]\nonumber\\ &&\times
           \frac{-g^{\beta\mu}+q^{\beta} q^{\mu}/m_{B^{\ast}}^{2}}{q^{2}-m^{2}_{B^{\ast}}}
           F^{2}\left(q^{2},m^2_{B^{\ast}}\right),\nonumber\\
 \mathcal{M}_{f}&=& \Big[g_{\Lambda^{0}_{b}pB^{\ast}}\bar{u}(p_{4})\gamma_{\beta}
         u(p_{2})\Big]\nonumber\\
           &&\times
          \Big[g_{B_{1}B^{\ast}\pi}(3ip_{1\mu}ip_{1\rho}
           -(ip^{\nu}_{1})(ip_{1\nu})g_{\mu\rho})\epsilon^{\mu}(p_{3})\Big]\nonumber\\ &&\times
           \frac{-g^{\beta\rho}+q^{\beta} q^{\rho}/m_{B^{\ast}}^{2}}{q^{2}-m^{2}_{B^{\ast}}}F^{2}
          \left(q^{2},m^2_{B^{\ast}}\right) ,\nonumber\\
\mathcal{M}_{g}&=&\Big[ig_{\Lambda^{0}_{b}pB}\bar{u}(p_{4})\gamma_{5}u(p_{2})\Big] \Big[g_{B^{*0}_{2}B\pi}\epsilon^{\alpha\beta}_{B^{\ast}_{2}}(ip_{1\alpha} ip_{1\beta})\Big]\nonumber\\
           &&\times \frac{1}{q^2-m_{B}^2} F^{2}\left(q^{2},m^2_{B}\right) ,\nonumber\\
\mathcal{M}_{h}&=&\Big[g_{\Lambda^{0}_{b}pB^{\ast}}\bar{u}(p_{4})
           \gamma _{\beta} u(p_{2}) \Big] ,\nonumber\\
           && \times \Big[-ig_{B^{\ast}_{2}B^{\ast}\pi}\varepsilon_{\alpha\lambda\eta\sigma}
\epsilon_{B^{\ast}_{2}}^{\zeta\eta}(iq^\sigma+ip_{3}^\sigma) (ip^{\zeta}_{1 } ip^{\lambda}_{1})\Big]\nonumber\\ &&\times \frac{-g^{\beta\alpha}+q^{\beta} q^{\alpha}/m_{B^{\ast}}^{2}}{q^{2}-m^{2}_{B^{\ast}}}F^{2} \left(q^{2},m^2_{B^{\ast}}\right).
\end{eqnarray}

Hereafter, we use $\mathcal{M}_{\mathcal{B}}$ refers to the amplitude for $\pi^- p \to \Lambda_b^0 \mathcal{B}$ with $\mathcal{B}=\{B, B^\ast, B_0, B_1^\prime, B_1, B_2\}$. Then the amplitudes for the concrete processes are,
\begin{eqnarray}
	\mathcal{M}_{B}&=& \mathcal{M}_a,\nonumber\\
	\mathcal{M}_{B^\ast}&=& \mathcal{M}_b+ \mathcal{M}_c,\nonumber\\
	\mathcal{M}_{B_0}&=& \mathcal{M}_d,\nonumber\\
    \mathcal{M}_{B_1^\prime}&=& \mathcal{M}_e,\nonumber\\
    \mathcal{M}_{B_1}&=& \mathcal{M}_f,\nonumber\\
    \mathcal{M}_{B_2}&=& \mathcal{M}_g+\mathcal{M}_h.
\end{eqnarray}

With the above preparations, we can calculate the cross sections for bottom meson productions in $\pi p$ scattering processes. The differential cross section can be,
\begin{eqnarray}
      \frac{d\sigma_\mathcal{B}}{d\cos\theta}=\frac{1}{32\pi s}\frac{| \vec{p}_{f}|}{\left|\vec{p}_{i}\right|} \left(\frac{1}{2}\overline{\left|\mathcal{M}_\mathcal{B}\right|^{2}}\right),
\end{eqnarray}
where $s=(p_{1}+p_{2})^2$ is the center of mass energy and $\theta$ is the angle between the outgoing bottom mesons and pion beam. $|\vec{p}_{i}|$ and $|\vec{p}_{f}|$ are the absolute values of the three momentum of the initial and final particles in the center-of-mass frame. $\mathcal{M}$ stands for amplitude, the factor $1/2$ is obtained by the spin average of initial states, and the overline indicates the sum over the spins of final states. In the same manner, we can obtain the cross sections for the bottom-strange meson productions in the $K^-p$ scattering processes.
\section{Numerical Results and discussions}
\label{Sec:Num}

In the amplitudes, a form factor in the monopole form is introduced to describe the internal structure and off-shell effect of the exchange mesons, which is~\cite{Chen:2013cpa,Chen:2014ccr},
\begin{eqnarray}
    F\left(q^{2},m^{2}_{B^{(\ast)}}\right)=
    \frac{m^{2}_{B^{(\ast)}}-\Lambda_{B^{(\ast)}}^{2}}{q^{2}
    -\Lambda^{2}_{B^{(\ast)}}}
    \label{Eq:FFs1}
\end{eqnarray}
  where $q$ and $m_{B^{(*)}}$ are the
  four-momentum and mass of the exchange bottom meson, respectively, while $\Lambda$ is a model parameter, which should not be far away from the mass of the exchanged meson \cite{Cheng:2004ru}. Thus, it is usually reparameterized as $\Lambda=m_{B^{*}}+\alpha \Lambda_{\mathrm{QCD}}$ with $\Lambda_{\mathrm{QCD}}=0.22$ GeV and $\alpha$ to be the model parameter, which is of order unity \cite{Tornqvist:1993vu,Tornqvist:1993ng,Locher:1993cc,Li:1996yn}. In the present work, we vary $\alpha$ from 0.6 to 1.4 to check the parameter dependences of our estimations.

\subsection{Cross Sections For $S-$ and $P-$Wave Bottom Meson Production}
\begin{figure}[htb]
  \includegraphics[width=8.5cm]{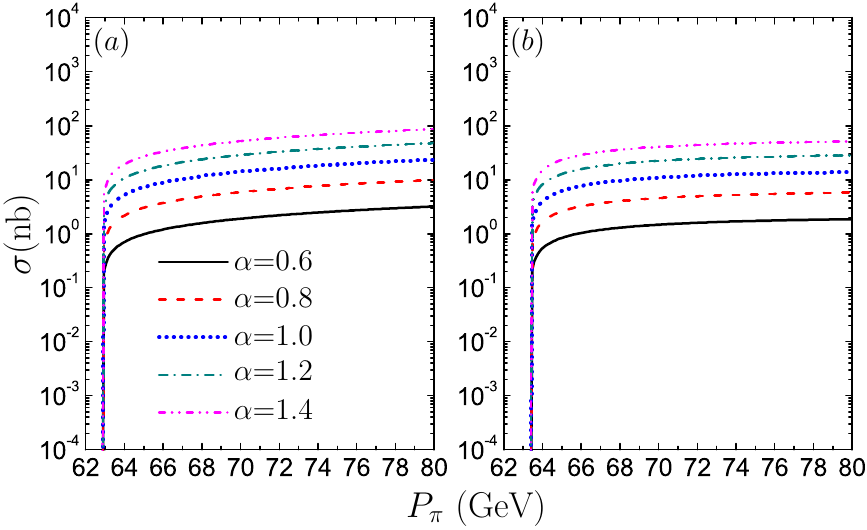}
   \caption{The cross sections for $\pi^- p \to B(5279) \Lambda_b^0$ (diagram (a)) and $\pi^- p \to B^\ast (5325) \Lambda_b^0$ (diagram (b)) depending on the momentum of the pion beam with typical $\alpha$ values, which are $0.6$, $0.8$, $1.0$, $1.2$ and $1.4$, respectively.  \label{Fig:CS-a} }
 \end{figure}

 \begin{figure}[htb]
\includegraphics[width=8.5cm]{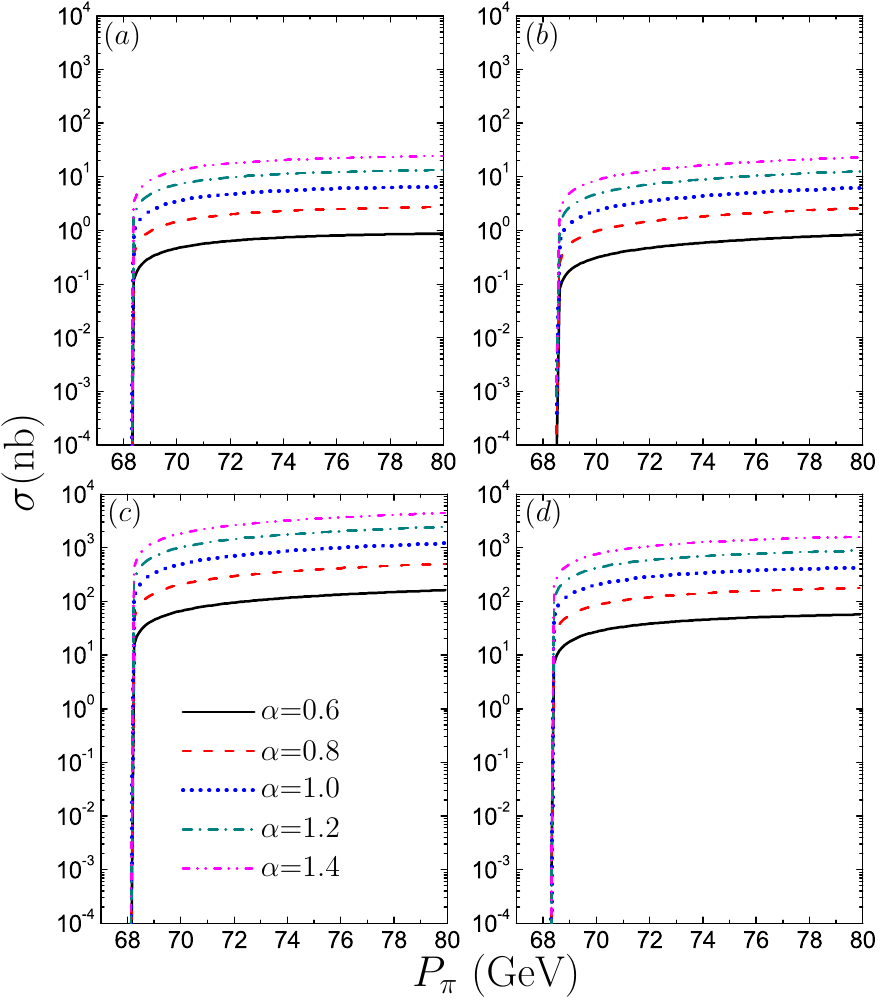}
 \caption{The same as Fig. \ref{Fig:CS-a}, but for $\pi^- p \to B_0^\ast (5738) \Lambda_b^0$ (diagram (a)),  $\pi^- p \to B_1^\prime (5757) \Lambda_b^0$ (diagram (b)), $\pi^- p \to B_1 (5721) \Lambda_b^0$ (diagram (c)) and $\pi^- p \to B_2^\ast (5747) \Lambda_b^0$ (diagram (d)).\label{Fig:CS-b}}

\end{figure}

In Fig.~\ref{Fig:CS-a}, the cross sections for the $S$-wave bottom meson production processes depending on the momentum of the incident pion beam are presented. Here, we take several typical $\alpha$ values, which are 0.6, 0.8, 1.0, 1.2 and 1.4, respectively, to check the model dependence of the cross sections. From Fig.~\ref{Fig:CS-a}-(a), we can find that the cross sections for $\pi^- p\to B(5279) \Lambda_b^0$ increase sharply near the threshold, which is $P_\pi=62.85$ GeV, then the cross sections become weakly dependent on the incident pion beam energy. However, the cross sections are still strongly dependent on the model parameter $\alpha$.  In particular, the cross sections vary from $3.19\sim 86.26$ nb at $P_\pi =80$ GeV with $\alpha$ increasing from 0.6 to 1.4. For $\pi p\to B^\ast (5325) \Lambda_b^0$, the $P_\pi$ dependences of the cross sections are very similar to those of $\pi^- p\to B(5279) \Lambda_b^0$, and at $P_\pi =80$ GeV, the cross sections are estimated to be $1.86\sim51.29 $ nb.

\begin{figure}[t]
\includegraphics[width=8.5cm]{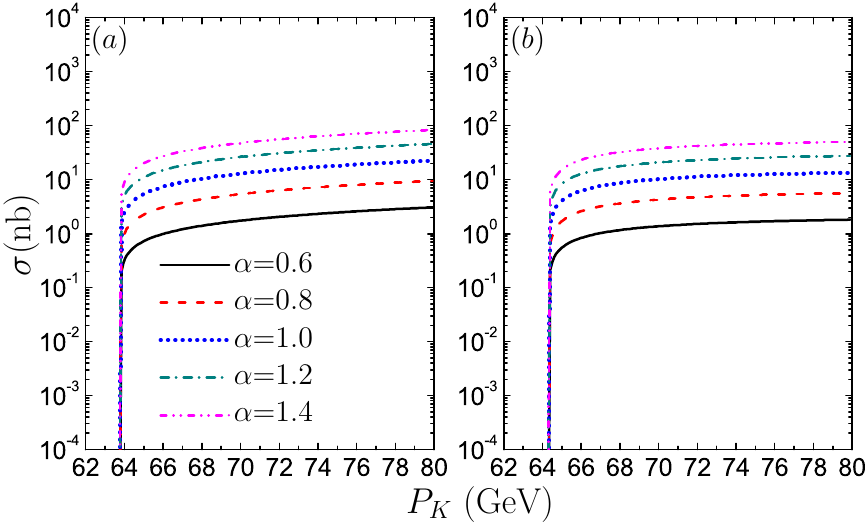}
\caption{The same as Fig.~\ref{Fig:CS-a}, but for $K^- p \to B_s(5366)\Lambda_b$ (diagram (a)) and $K^- p \to B_s^{\ast}(5415) $ (diagram (b)). \label{Fig:CS-c} }
\end{figure}
\begin{figure}[t]
\includegraphics[width=8.5cm]{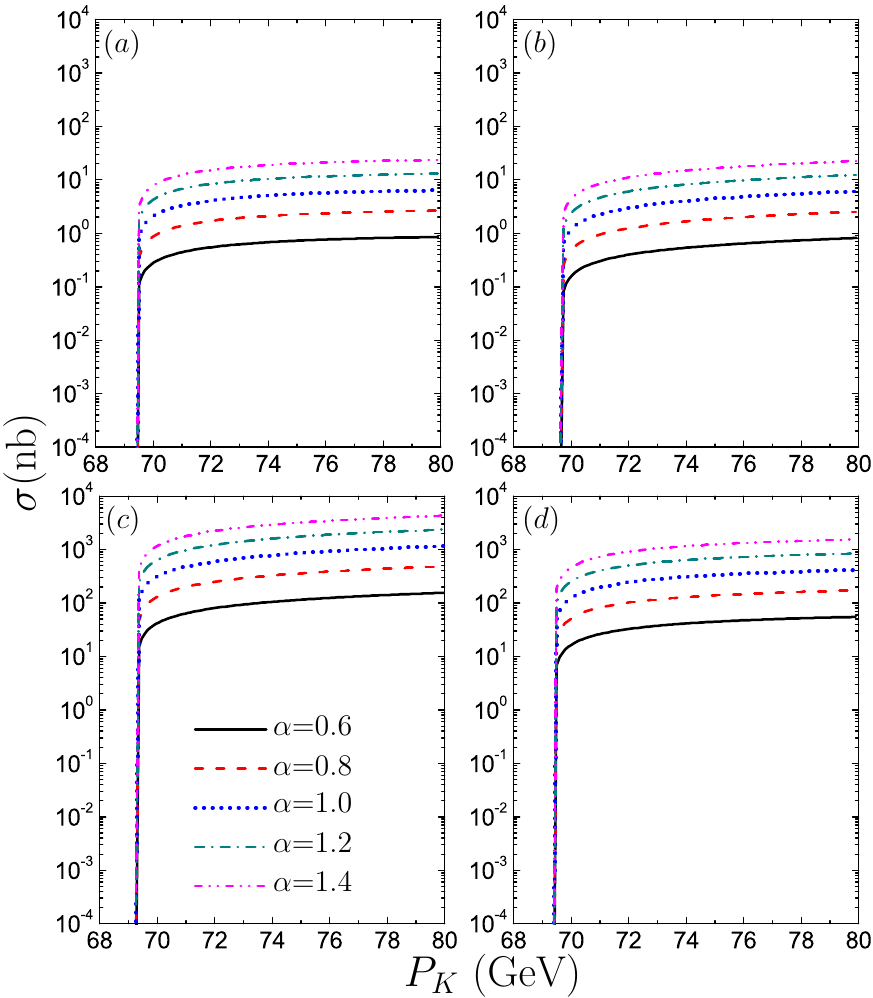}
\caption{The same as Fig.~\ref{Fig:CS-a}, but for $K^- p \to B_{s0}^\ast (5841) \Lambda_b^0$ (diagram (a)),  $K^- p \to B_{s1}^\prime (5859) \Lambda_b^0$ (diagram (b)), $K^- p \to B_{s1} (5830) \Lambda_b^0$ (diagram (c)) and $K^- p \to B_{s2}^\ast (5840) \Lambda_b^0$ (diagram (d)). \label{Fig:CS-d} }
\end{figure}

Our estimated cross sections for the $P$ wave bottom meson productions in the $\pi p$ scattering are presented in Fig.~\ref{Fig:CS-b}. The $P_\pi$ dependences of the cross sections are very similar to those of the $S$-wave bottom meson production processes but in different magnitudes. In particular, at the incident beam energy $p_\pi=80$ GeV, the cross sections for $B_0^\ast(5738)$, $B_1^\prime (5757)$, $B_1(5275)$ and $B_2^\ast (5747)$ are estimated to be $0.87\sim24.25$ nb, $0.84\sim23.14$ nb, $162.35\sim4477.66$ nb, and $57.16 \sim1604.43$ nb, respectively. From the figure, we can find that the estimated cross sections for the bottom mesons in the same doublet are similar, which is consistent with the degeneration of the states in the same doublet in the heavy-quark limit. However, the cross sections for the bottom mesons in the $T$ doublet are about two orders larger than those for the bottom mesons in the $S$ doublet.

\begin{figure}[t]
\includegraphics[width=8cm]{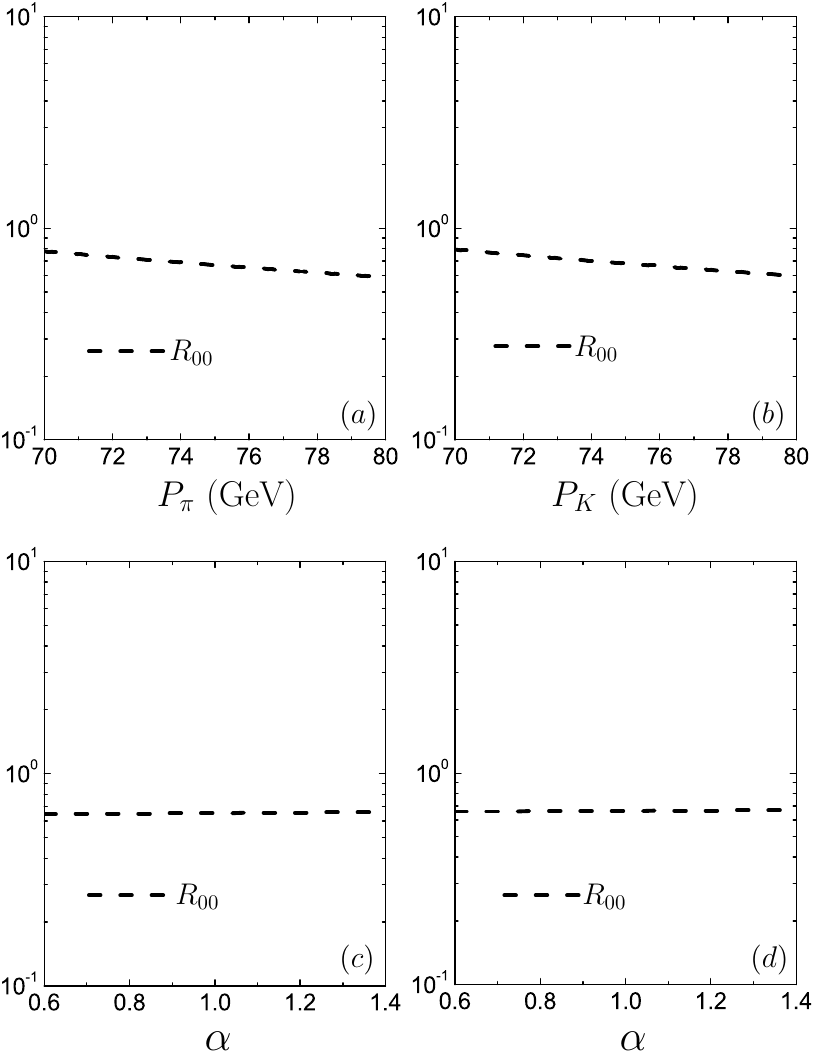}
\caption{The incident beam energy and model parameter $\alpha$ dependences of $R_{00}$ for bottom meson productions (diagrams (a) and (c)) and bottom-strange meson productions (diagrams (b) and (d)).
  \label{Fig:SwaveRatio1} }
\end{figure}

\begin{figure}[htb]
\includegraphics[width=8.5cm]{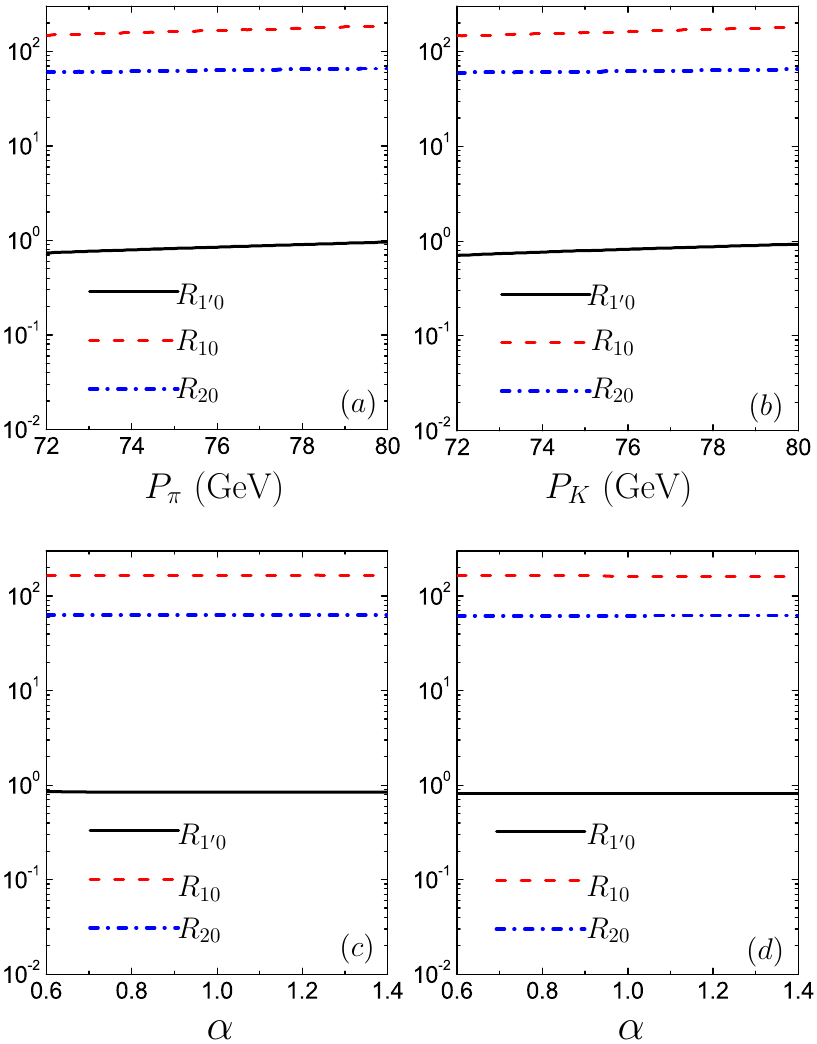}
\caption{The same as Fig.~\ref{Fig:SwaveRatio1}, but for P-wave bottom and bottom-strange meson productions.  \label{Fig:PwaveRatio1} }
\end{figure}

\subsection{Cross Sections For $S-$ and $P-$wave bottom-strange meson production}

In a very similar manner, we can estimate the  $S-$ and $P-$wave bottom-strange meson productions in the kaon induced reactions on proton targets. The cross sections for the $S$-wave and $P$-wave bottom-strange meson productions are presented in Figs.~\ref{Fig:CS-c} and \ref{Fig:CS-d}, respectively. The estimated $p_K$ dependences of the cross sections for the $S$-wave bottom-strange mesons are very similar to those for bottom meson productions as shown in Fig.~\ref{Fig:CS-a}. For the process $Kp\to B_s(5366) \Lambda_b$, the cross sections are estimated to be $3.07\sim 82.98$ nb at $P_K$=80 GeV with $\alpha$ increasing from 0.6 to 1.4. While cross sections for $Kp\to B_s^\ast(5415) \Lambda_b$ in Fig.~\ref{Fig:CS-c}-(b) can reach up to 50 nb at $P_K$=80 GeV with $\alpha$ =1.4.

In addition, in Fig.~\ref{Fig:CS-d}, we present the cross sections for $P$- wave bottom-strange mesons production processes. From the figure, one can find that at $P_K=80$ GeV, the cross sections are estimated to be $0.86\sim23.77$ nb, $0.82\sim 22.66$ nb, $155.74\sim 4302.51$ nb, and $55.01\sim 1544.13$ nb for $B_{s0}^\ast(5841)$, $B_{s1}^\prime(5859)$, $B_{s1}(5830)$, and $B_{s2}^\ast(5840)$,  respectively. The cross sections for the productions of the bottom-strange meson in the same doublet are similar, which is in line with the conclusion for bottom mesons. Overall, our estimations indicate that the cross sections for the bottom-strange meson productions are similar to those for the bottom meson productions, which reflects the SU(3) symmetry for the $u$, $d$ and $s$ quarks.

\subsection{Cross Sections Ratio}

From Figs.~\ref{Fig:CS-a}-\ref{Fig:CS-d}, one can find that the momentum and the model parameter dependences of the estimated cross sections are very similar due to the heavy-quark limit. In order to further investigate the degenerate properties of the cross sections, we define the ratios of the cross sections as,
\begin{eqnarray}
R_{00} &=& \frac{\sigma(\pi^- p \to B^{\ast}(5325) \Lambda^{0}_b)}{\sigma(\pi^- p \to B(5279) \Lambda^{0}_b)}, \nonumber \\
R_{1^\prime 0} &=& \frac{\sigma(\pi p \to B_1^{\prime}(5757) \Lambda^{0}_b)}
               {\sigma(\pi^- p \to B_0^{\ast}(5738)\Lambda^{0}_b)}, \nonumber
 \end{eqnarray}
 \begin{eqnarray}
 R_{1 0} &=& \frac{\sigma(\pi^- p \to B_1(5721)^{0} \Lambda^{0}_b)}{\sigma(\pi^- p \to B_0^{\ast} (5738)\Lambda^{0}_b)}, \nonumber \\
R_{2 0} &=& \frac{\sigma(\pi^- p \to B_2^\ast(5747)^{0} \Lambda^{0}_b)}{\sigma(\pi^- p \to B_0^{\ast}(5738)\Lambda^{0}_b)}.              
 \label{Eq:cross ratio}
\end{eqnarray}
Similarly, we can define the ratios of cross sections for bottom-strange meson production processes.

The estimated incident beam energy and the model parameter $\alpha$ dependences of $R_{00}$ for bottom meson productions and bottom-strange meson productions are presented in Fig.~\ref{Fig:SwaveRatio1}. For the bottom meson production processes, the ratio $R_{00}$ is estimated to be $0.59 \sim 0.78$ with $70\ \rm GeV <P_\pi < 80\ \rm GeV$ and $\alpha=1.4$, which indicate that the ratio is not strongly dependent on the incident beam momentum. In addition, from Fig.~\ref{Fig:SwaveRatio1}-(c), one can find that the ratio is almost independent on the model parameter $\alpha$ in a certain $P_\pi$. In particular, the ratio $R_{00}$ is estimated to be $ 0.65 \sim 0.67 $ with $0.6 <\alpha<1.4$ and $P_\pi =76$ GeV. Similar conclusions can be drawn for the bottom-strange meson production processes as shown in Figs.~\ref{Fig:SwaveRatio1}-(b) and (d). The ratio is estimated to be $0.60 \sim 0.80$ with $70\ \rm GeV <P_K < 80\ \rm GeV$ and $\alpha=1.4$, while it is evaluated to be $0.66 \sim 0.67$ with $0.6 <\alpha<1.4$ and $P_K =76$ GeV. The ratios $R_{00}$ for both the bottom and bottom-strange meson production processes are of order one, indicating the degenerate wave singlet and triplet states.

For the $P$-wave bottom and bottom-strange meson production processes, their cross sections ratios are presented in Fig.~\ref{Fig:PwaveRatio1}. From Fig.~\ref{Fig:PwaveRatio1}-(a) and Fig.~\ref{Fig:PwaveRatio1}-(b), one can find that with $\alpha=1.4$, the ratio $R_{1^\prime 0}$ is estimated to be $0.74 \sim 0.96$ and $0.71 \sim 0.93 $ for the bottom meson production and bottom-strange meson production processes, respectively. Moreover, for bottom meson production processes, $R_{10}$ and $R_{20}$ are determined to be $148.93 \sim 185.60$ and $60.25 \sim 65.91$, respectively, while for the bottom-strange meson production processes, these two ratios are estimated to be $146.48\sim 181.45$ and $59.70\sim 64.68$, respectively. As for the $\alpha$ dependences, our estimations show that these ratios are very weakly dependent on the model parameter $\alpha$. Thus, these ratios could serve as a good test for the present estimations.

\subsection{Differential Cross Sections}

\begin{figure}[t]
\includegraphics[width=8.5cm]{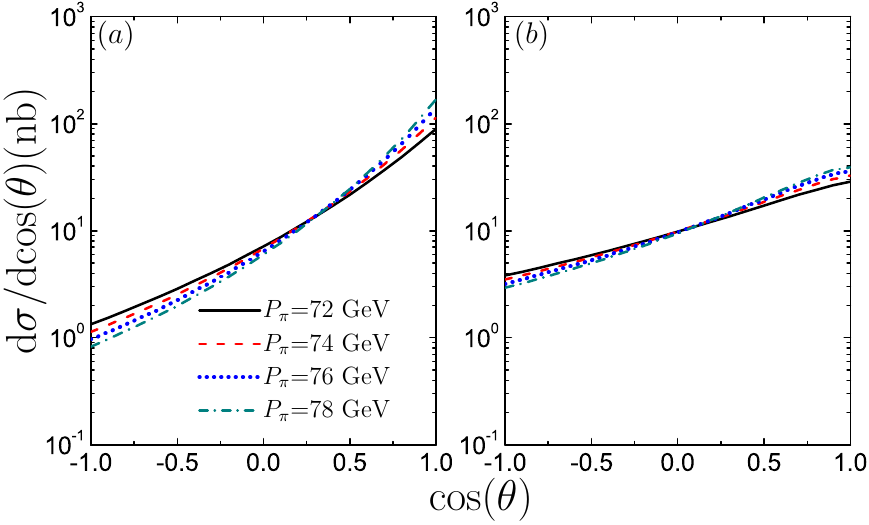}
  \caption{ The same as Fig.~\ref{Fig:CS-a}, but for differential cross sections depending on cos$\theta$. \label{Fig:B-e} }
\end{figure}

\begin{figure}[t]
 \includegraphics[width=8.5cm]{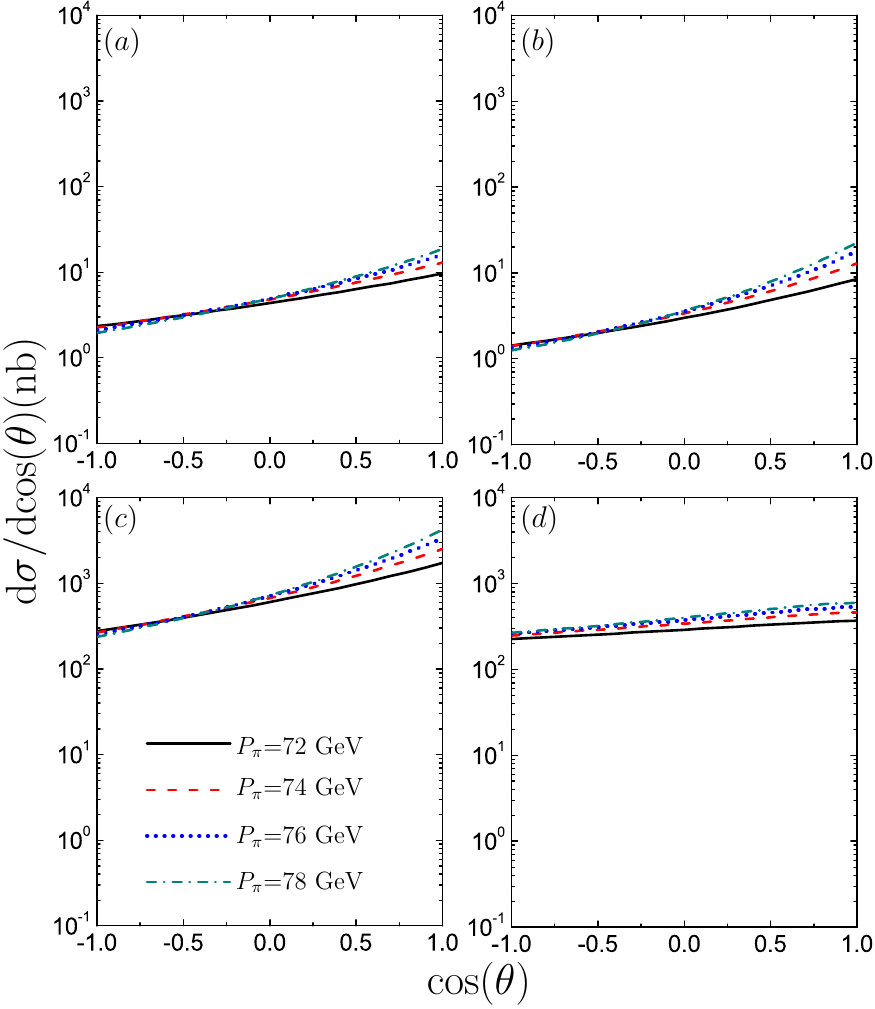}
  \caption{The same as Fig.~\ref{Fig:CS-b}, but for differential cross sections depending on cos$\theta$. \label{Fig:B-f} }
\end{figure}

\begin{figure}[t]
\includegraphics[width=8.5cm]{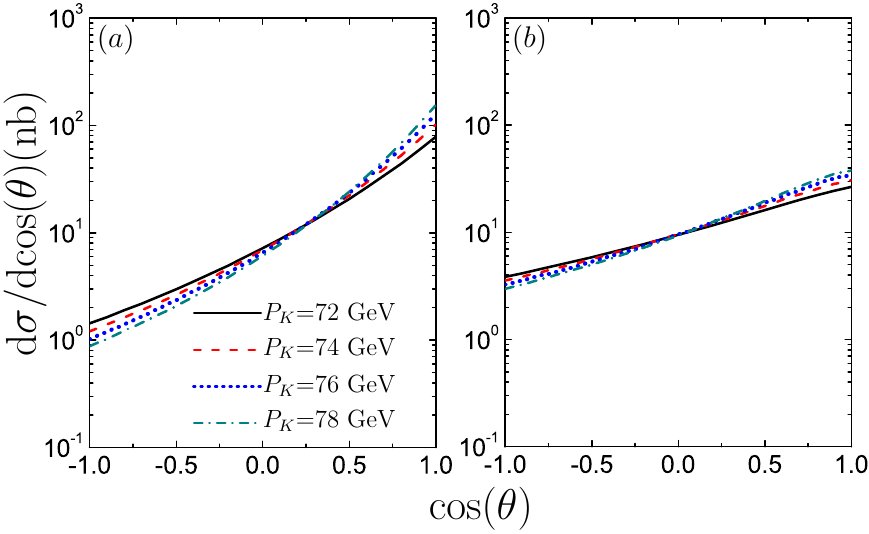}
\caption{The same as Fig.~\ref{Fig:CS-c}, but for differential cross sections depending on cos$\theta$. .\label{Fig:BS-g} }
\end{figure}

\begin{figure}[t]
\includegraphics[width=8.7cm]{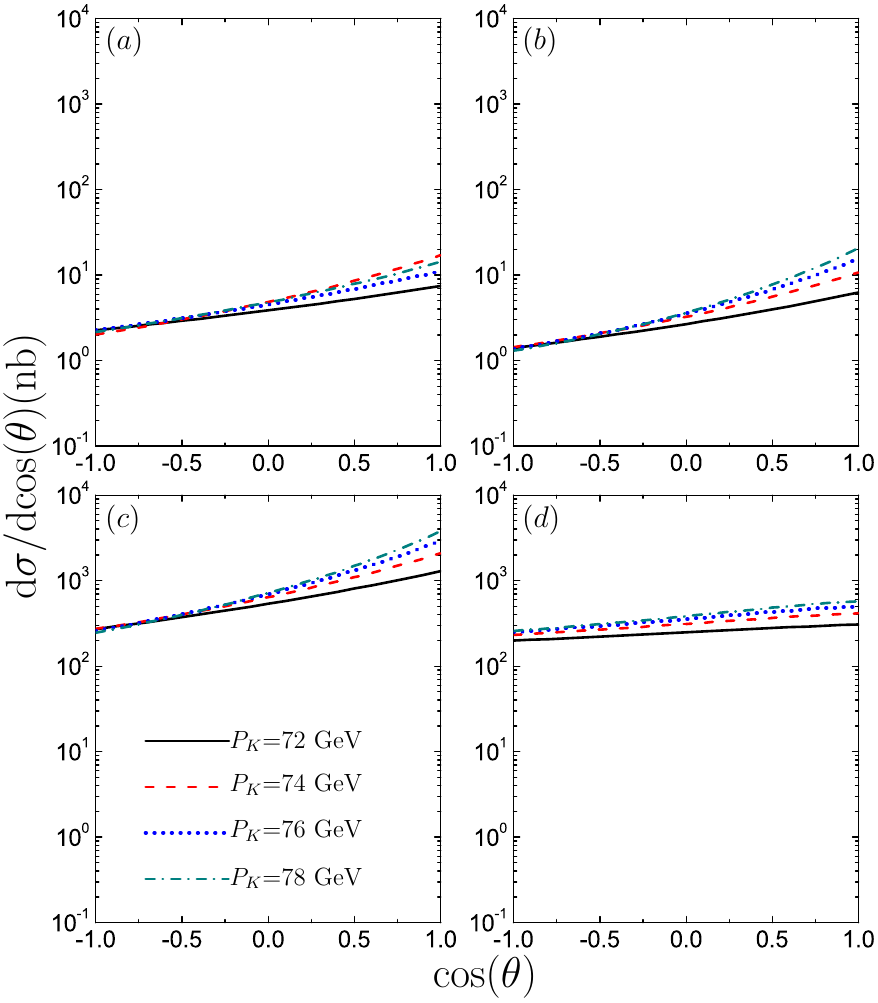}
\caption{The same as Fig.~\ref{Fig:CS-d}, but for differential cross sections depending on cos$\theta$. \label{Fig:BS-h} }
\end{figure}

In addition to the total cross sections and cross section ratios, we also calculated the differential cross sections for the $S$-, $P$-wave bottom and bottom- stranged mesons production processes depending on $\cos\theta$, where $\theta$ is the angle between the outgoing bottom or bottom-strange meson and the incident beam direction. Our estimates indicate that the shape of differential cross sections are weakly dependent on the model parameter, therefore in the following discussion, we fix the model parameter $\alpha = 1.0$. Several typical incident beam energies, which are $72,\ 74,\ 76,\ 78$ GeV, respectively, are considered in the present estimations.

In Figs.~\ref{Fig:B-e} and \ref{Fig:B-f} , we present the differential cross sections for $S$- and $P$-wave bottom mesons production processes, respectively. One of the main remarkable features of these figures is that the differential cross sections are mainly concentrated in the forward angle limit. We also notice that for the $P$-wave bottom mesons production processes with $P_\pi=72$ GeV, the differential cross sections are weakly dependent on $\cos \theta$, especially for $\pi p\to B_2^\ast (5747) \Lambda_b^0$ process as shown in Fig.~\ref{Fig:B-f}-(d). The reason is that $P_\pi=72$ GeV is a bit close to the threshold of the process $\pi p\to B_2^\ast (5747) \Lambda_b^0$, and the momentum of the outgoing $B_2^\ast (5747)$ or $\Lambda_b^0$ is rather small due to the large mass of these particles; thus, in this case, the differential cross sections become weakly dependent on $\cos\theta $. As for the differential cross sections for the $S$- ,$P$-wave bottom-strange mesons production process, they exhibit similar $\cos \theta $ dependences as shown in Figs.~\ref{Fig:BS-g} and \ref{Fig:BS-h}.

\section{Summary}
In the present work, we propose to explore the production of the bottom and bottom-strange mesons in the high-energy pion- and kaon-induced reactions on a proton target. The $S$- and $P$-wave bottom and bottom-strange mesons production processes are investigated with an effective Lagrangian approach. The total cross sections and differential cross sections of the relevant processes are calculated. Our estimations indicate that the cross sections for the states in the $H$ doublet are similar, ranging from $1$ nb to several tens of nanobarns, with $\alpha$ increasing from 0.6 to 1.4 and the incident beam energy to be 80 GeV. For the $P$ wave bottom and bottom-strange mesons production processes, we notice that the cross sections for the states in the $S$ doublet are similar, while the cross sections for the states in the $T$ doublet are also in the same order. This property is consistent with the expectation of a heavy-quark limit. Fro the differential cross sections, we find that the differential cross sections are mainly concentrated in the forward angle limit.

In addition, we find that the model parameter dependences of the cross sections for the relevant processes are very similar, thus, we further define the cross sections ratios to evade the uncertainties caused by the model parameter. We find that these ratios are rather weakly dependent on parameter $\alpha$ as we expected. In the consider incident beam energy range, $R_{00}$ is estimated to be $0.66\sim 0.67 $ and $0.65 \sim 0.66 $ for the bottom and bottom-strange mesons production processes. For the $P$ bottom mesons production processes, $R_{1^\prime 0}$, $R_{10}$ and $R_{20}$ are estimated to be $0.74\sim0.96 $, $148.93\sim185.61 $ and $ 60.26\sim65.91 $, respectively, while for the bottom-strange mesons production processes, these ratios are evaluated to be $0.71\sim0.93$, $146.48\sim181.45$ and $59.81\sim 64.68$, respectively.

\section* {Acknowledgements}
This work is supported by the National Natural Science Foundation of China under Grant Nos. 12175037,  and 12335001.


\begin{thebibliography}{00}

\bibitem{E288:1977xhf}
S.~W.~Herb \textit{et al.} [E288],
Phys. Rev. Lett. \textbf{39} (1977), 252-255
doi:10.1103/PhysRevLett.39.252




\bibitem{Pluto:1978tuc}
C.~Berger \textit{et al.} [Pluto],
Phys. Lett. B \textbf{76} (1978), 243-245
doi:10.1016/0370-2693(78)90287-3


\bibitem{CLEO:1983mma}
S.~Behrends \textit{et al.} [CLEO],
Phys. Rev. Lett. \textbf{50}, 881-884 (1983)
doi:10.1103/PhysRevLett.50.881


\bibitem{ALEPH:1993mpb}
D.~Buskulic \textit{et al.} [ALEPH],
Phys. Lett. B \textbf{311}, 425-430 (1993)
[erratum: Phys. Lett. B \textbf{316}, 631 (1993)]
doi:10.1016/0370-2693(93)90588-9



\bibitem{CDF:1993pzh}
F.~Abe \textit{et al.} [CDF],
Phys. Rev. Lett. \textbf{71}, 1685-1689 (1993)
doi:10.1103/PhysRevLett.71.1685


\bibitem{ParticleDataGroup:2022pth}
R.~L.~Workman \textit{et al.} [Particle Data Group],
PTEP \textbf{2022} (2022), 083C01
doi:10.1093/ptep/ptac097

\bibitem{OPAL:1994hqv}
R.~Akers \textit{et al.} [OPAL],
Z. Phys. C \textbf{66} (1995), 19-30
doi:10.1007/BF01496577



\bibitem{DELPHI:1994fnu}
P.~Abreu \textit{et al.} [DELPHI],
Phys. Lett. B \textbf{345} (1995), 598-608
doi:10.1016/0370-2693(94)01696-A


\bibitem{ALEPH:1995ikc}
D.~Buskulic \textit{et al.} [ALEPH],
Z. Phys. C \textbf{69} (1996), 393-404
doi:10.1007/BF02907419


\bibitem{ALEPH:1998unp}
R.~Barate \textit{et al.} [ALEPH],
Phys. Lett. B \textbf{425} (1998), 215-226
doi:10.1016/S0370-2693(98)00180-4

\bibitem{L3:1999pdo}
M.~Acciarri \textit{et al.} [L3],
Phys. Lett. B \textbf{465}, 323-334 (1999)
doi:10.1016/S0370-2693(99)01067-9
[arXiv:hep-ex/9909018 [hep-ex]].

\bibitem{CDF:1999zui}
T.~Affolder \textit{et al.} [CDF],
Phys. Rev. D \textbf{64}, 072002 (2001)
doi:10.1103/PhysRevD.64.072002


\bibitem{D0:2007vzd}
V.~M.~Abazov \textit{et al.} [D0],
Phys. Rev. Lett. \textbf{99} (2007), 172001
doi:10.1103/PhysRevLett.99.172001
[arXiv:0705.3229 [hep-ex]].


\bibitem{CDF:2008qzb}
T.~Aaltonen \textit{et al.} [CDF],
Phys. Rev. Lett. \textbf{102}, 102003 (2009)
doi:10.1103/PhysRevLett.102.102003
[arXiv:0809.5007 [hep-ex]].


\bibitem{CDF:2013www}
T.~A.~Aaltonen \textit{et al.} [CDF],
Phys. Rev. D \textbf{90} (2014) no.1, 012013
doi:10.1103/PhysRevD.90.012013
[arXiv:1309.5961 [hep-ex]].

\bibitem{LHCb:2015aaf}
R.~Aaij \textit{et al.} [LHCb],
JHEP \textbf{04}, 024 (2015)
doi:10.1007/JHEP04(2015)024
[arXiv:1502.02638 [hep-ex]].








\bibitem{CDF:2007avt}
T.~Aaltonen \textit{et al.} [CDF],
Phys. Rev. Lett. \textbf{100}, 082001 (2008)
doi:10.1103/PhysRevLett.100.082001
[arXiv:0710.4199 [hep-ex]].


\bibitem{D0:2007die}
V.~M.~Abazov \textit{et al.} [D0],
Phys. Rev. Lett. \textbf{100}, 082002 (2008)
doi:10.1103/PhysRevLett.100.082002
[arXiv:0711.0319 [hep-ex]].


\bibitem{LHCb:2012iuq}
R.~Aaij \textit{et al.} [LHCb],
Phys. Rev. Lett. \textbf{110}, no.15, 151803 (2013)
doi:10.1103/PhysRevLett.110.151803
[arXiv:1211.5994 [hep-ex]].



\bibitem{CMS:2018wcx}
A.~M.~Sirunyan \textit{et al.} [CMS],
Eur. Phys. J. C \textbf{78}, no.11, 939 (2018)
doi:10.1140/epjc/s10052-018-6390-z
[arXiv:1809.03578 [hep-ex]].



\bibitem{LHCb:2020pet}
R.~Aaij \textit{et al.} [LHCb],
Eur. Phys. J. C \textbf{81}, no.7, 601 (2021)
doi:10.1140/epjc/s10052-021-09305-3
[arXiv:2010.15931 [hep-ex]].


\bibitem{Isgur:1998kr}
N.~Isgur,
Phys. Rev. D \textbf{57}, 4041-4053 (1998)
doi:10.1103/PhysRevD.57.4041

\bibitem{Asghar:2018tha}
I.~Asghar, B.~Masud, E.~S.~Swanson, F.~Akram and M.~Atif Sultan,
Eur. Phys. J. A \textbf{54} (2018) no.7, 127
doi:10.1140/epja/i2018-12558-6
[arXiv:1804.08802 [hep-ph]].


\bibitem{Ebert:1997nk}
D.~Ebert, V.~O.~Galkin and R.~N.~Faustov,
Phys. Rev. D \textbf{57} (1998), 5663-5669
[erratum: Phys. Rev. D \textbf{59} (1999), 019902]
doi:10.1103/PhysRevD.59.019902
[arXiv:hep-ph/9712318 [hep-ph]].


\bibitem{Liu:2013maa}
J.~B.~Liu and M.~Z.~Yang,
JHEP \textbf{07} (2014), 106
doi:10.1007/JHEP07(2014)106
[arXiv:1307.4636 [hep-ph]].

\bibitem{Liu:2015lka}
J.~B.~Liu and M.~Z.~Yang,
Phys. Rev. D \textbf{91} (2015) no.9, 094004
doi:10.1103/PhysRevD.91.094004
[arXiv:1501.04266 [hep-ph]].

\bibitem{Wang:2016itc}
W.~Wang, D.~Jia and D.~Chen,
Nucl. Phys. Rev. \textbf{33} (2016) no.1, 30-35
doi:10.11804/NuclPhysRev.33.01.030


\bibitem{Orsland:1998de}
A.~H.~Orsland and H.~Hogaasen,
Eur. Phys. J. C \textbf{9}, 503-510 (1999)
doi:10.1007/s100529900042
[arXiv:hep-ph/9812347 [hep-ph]].

\bibitem{Matsuki:2006zoi}
T.~Matsuki, T.~Morii and K.~Sudoh,
Prog. Theor. Phys. \textbf{117}, 1077-1098 (2007)
doi:10.1143/PTP.117.1077
[arXiv:hep-ph/0605019 [hep-ph]].


\bibitem{DiPierro:2001dwf}
M.~Di Pierro and E.~Eichten,
Phys. Rev. D \textbf{64}, 114004 (2001)
doi:10.1103/PhysRevD.64.114004
[arXiv:hep-ph/0104208 [hep-ph]].



\bibitem{Xiao:2014ura}
L.~Y.~Xiao and X.~H.~Zhong,
Phys. Rev. D \textbf{90}, no.7, 074029 (2014)
doi:10.1103/PhysRevD.90.074029
[arXiv:1407.7408 [hep-ph]].


\bibitem{Sun:2014wea}
Y.~Sun, Q.~T.~Song, D.~Y.~Chen, X.~Liu and S.~L.~Zhu,
Phys. Rev. D \textbf{89}, no.5, 054026 (2014)
doi:10.1103/PhysRevD.89.054026
[arXiv:1401.1595 [hep-ph]].


\bibitem{Falk:1995th}
A.~F.~Falk and T.~Mehen,
Phys. Rev. D \textbf{53} (1996), 231-240
doi:10.1103/PhysRevD.53.231
[arXiv:hep-ph/9507311 [hep-ph]].



\bibitem{Eichten:1993ub}
E.~J.~Eichten, C.~T.~Hill and C.~Quigg,
Phys. Rev. Lett. \textbf{71}, 4116-4119 (1993)
doi:10.1103/PhysRevLett.71.4116
[arXiv:hep-ph/9308337 [hep-ph]].

\bibitem{Lewis:2000sv}
R.~Lewis and R.~M.~Woloshyn,
Phys. Rev. D \textbf{62}, 114507 (2000)
doi:10.1103/PhysRevD.62.114507
[arXiv:hep-lat/0003011 [hep-lat]].



\bibitem{Colangelo:2006kx}
P.~Colangelo, F.~De Fazio and R.~Ferrandes,
Nucl. Phys. B Proc. Suppl. \textbf{163}, 177-182 (2007)
doi:10.1016/j.nuclphysbps.2006.09.005
[arXiv:hep-ph/0609072 [hep-ph]].



\bibitem{Colangelo:2012xi}
P.~Colangelo, F.~De Fazio, F.~Giannuzzi and S.~Nicotri,
Phys. Rev. D \textbf{86}, 054024 (2012)
doi:10.1103/PhysRevD.86.054024
[arXiv:1207.6940 [hep-ph]].

\bibitem{Wang:2014cta}
Z.~G.~Wang,
Eur. Phys. J. Plus \textbf{129}, 186 (2014)
doi:10.1140/epjp/i2014-14186-y
[arXiv:1401.7580 [hep-ph]].


\bibitem{Green:2003zza}
A.~M.~Green \textit{et al.} [UKQCD],
Phys. Rev. D \textbf{69}, 094505 (2004)
doi:10.1103/PhysRevD.69.094505
[arXiv:hep-lat/0312007 [hep-lat]].



\bibitem{Lahde:1999ih}
T.~A.~Lahde, C.~J.~Nyfalt and D.~O.~Riska,
Nucl. Phys. A \textbf{674} (2000), 141-167
doi:10.1016/S0375-9474(00)00154-8
[arXiv:hep-ph/9908485 [hep-ph]].



\bibitem{BaBar:2003cdx}
B.~Aubert \textit{et al.} [BaBar],
Phys. Rev. D \textbf{69} (2004), 031101
doi:10.1103/PhysRevD.69.031101
[arXiv:hep-ex/0310050 [hep-ex]].


\bibitem{SELEX:2004drx}
A.~V.~Evdokimov \textit{et al.} [SELEX],
Phys. Rev. Lett. \textbf{93} (2004), 242001
doi:10.1103/PhysRevLett.93.242001
[arXiv:hep-ex/0406045 [hep-ex]].

\bibitem{CLEO:2003ggt}
D.~Besson \textit{et al.} [CLEO],
Phys. Rev. D \textbf{68} (2003), 032002
[erratum: Phys. Rev. D \textbf{75} (2007), 119908]
doi:10.1103/PhysRevD.68.032002
[arXiv:hep-ex/0305100 [hep-ex]].

\bibitem{Belle:2003kup}
Y.~Mikami \textit{et al.} [Belle],
Phys. Rev. Lett. \textbf{92} (2004), 012002
doi:10.1103/PhysRevLett.92.012002
[arXiv:hep-ex/0307052 [hep-ex]].


\bibitem{Belle:2003guh}
P.~Krokovny \textit{et al.} [Belle],
Phys. Rev. Lett. \textbf{91} (2003), 262002
doi:10.1103/PhysRevLett.91.262002
[arXiv:hep-ex/0308019 [hep-ex]].

\bibitem{BaBar:2004yux}
B.~Aubert \textit{et al.} [BaBar],
Phys. Rev. Lett. \textbf{93} (2004), 181801
doi:10.1103/PhysRevLett.93.181801
[arXiv:hep-ex/0408041 [hep-ex]].

\bibitem{BaBar:2006eep}
B.~Aubert \textit{et al.} [BaBar],
Phys. Rev. D \textbf{74} (2006), 032007
doi:10.1103/PhysRevD.74.032007
[arXiv:hep-ex/0604030 [hep-ex]].

\bibitem{BaBar:2003oey}
B.~Aubert \textit{et al.} [BaBar],
Phys. Rev. Lett. \textbf{90} (2003), 242001
doi:10.1103/PhysRevLett.90.242001
[arXiv:hep-ex/0304021 [hep-ex]].

\bibitem{Datta:2003re}
  A.~Datta and P.~J.~O'donnell,
  Understanding the nature of $D_s(2317)$ and $D_s(2460)$ through nonleptonic B decays,
  Phys.\ Lett.\ B {\bf 572} , 164 (2003).

\bibitem{LHCb:2013jjb}
R.~Aaij \textit{et al.} [LHCb],
JHEP \textbf{09}, 145 (2013)
doi:10.1007/JHEP09(2013)145
[arXiv:1307.4556 [hep-ex]].

\bibitem{FOCUS:2003gru}
J.~M.~Link \textit{et al.} [FOCUS],
Phys. Lett. B \textbf{586}, 11-20 (2004)
doi:10.1016/j.physletb.2004.02.017
[arXiv:hep-ex/0312060 [hep-ex]].



\bibitem{ZEUS:2012gyr}
H.~Abramowicz \textit{et al.} [ZEUS],
Nucl. Phys. B \textbf{866}, 229-254 (2013)
doi:10.1016/j.nuclphysb.2012.09.007
[arXiv:1208.4468 [hep-ex]].


\bibitem{ACCMOR:1990xso}
S.~Barlag \textit{et al.} [ACCMOR],
Z. Phys. C \textbf{46}, 563-568 (1990)
doi:10.1007/BF01560257



\bibitem{BigBubbleChamberNeutrino:1995amd}
A.~E.~Asratvan \textit{et al.} [Big Bubble Chamber Neutrino],
Z. Phys. C \textbf{68}, 43-46 (1995)
doi:10.1007/BF01579803


\bibitem{Nagae:2008zz}
T.~Nagae,
Nucl. Phys. A \textbf{805} (2008), 486-493
doi:10.1016/j.nuclphysa.2008.02.287


\bibitem{Obraztsov:2016lhp}
V.~Obraztsov [OKA],
Nucl. Part. Phys. Proc. \textbf{273-275} (2016), 1330-1333
doi:10.1016/j.nuclphysbps.2015.09.213

\bibitem{Nerling:2012er}
F.~Nerling [COMPASS],
EPJ Web Conf. \textbf{37} (2012), 01016
doi:10.1051/epjconf/20123701016
[arXiv:1208.0487 [hep-ex]].


\bibitem{Wise:1992hn}
M.~B.~Wise,
Phys. Rev. D \textbf{45} (1992) no.7, R2188
doi:10.1103/PhysRevD.45.R2188

\bibitem{Quintans:2022utc}
C.~Quintans [AMBER],
Few Body Syst. \textbf{63} (2022) no.4, 72
doi:10.1007/s00601-022-01769-7



\bibitem{Ebert:2001zm}
D.~Ebert, R.~N.~Faustov and V.~O.~Galkin,
AIP Conf. Proc. \textbf{619} (2002) no.1, 336-345
doi:10.1063/1.1482462
[arXiv:hep-ph/0110190 [hep-ph]].


\bibitem{He:2016pfa}
J.~He,
Phys. Rev. D \textbf{95} (2017) no.7, 074004
doi:10.1103/PhysRevD.95.074004
[arXiv:1607.03223 [hep-ph]].


\bibitem{Dong:2014ksa}
Y.~Dong, A.~Faessler, T.~Gutsche and V.~E.~Lyubovitskij,
Phys. Rev. D \textbf{90} (2014) no.9, 094001
doi:10.1103/PhysRevD.90.094001
[arXiv:1407.3949 [hep-ph]].




\bibitem{Chen:2013cpa}
D.~Y.~Chen, X.~Liu and T.~Matsuki,
Phys. Rev. D \textbf{87} (2013) no.9, 094010
doi:10.1103/PhysRevD.87.094010
[arXiv:1304.0372 [hep-ph]].


\bibitem{Chen:2014ccr}
D.~Y.~Chen, X.~Liu and T.~Matsuki,
Phys. Rev. D \textbf{90} (2014) no.3, 034019
doi:10.1103/PhysRevD.90.034019
[arXiv:1406.6763 [hep-ph]].



\bibitem{Cheng:2004ru}
H.~Y.~Cheng, C.~K.~Chua and A.~Soni,
Phys. Rev. D \textbf{71} (2005), 014030
doi:10.1103/PhysRevD.71.014030
[arXiv:hep-ph/0409317 [hep-ph]].



\bibitem{Tornqvist:1993vu}
N.~A.~Tornqvist,
Nuovo Cim. A \textbf{107} (1994), 2471-2476
doi:10.1007/BF02734018
[arXiv:hep-ph/9310225 [hep-ph]].





\bibitem{Tornqvist:1993ng}
N.~A.~Tornqvist,
Z. Phys. C \textbf{61} (1994), 525-537
doi:10.1007/BF01413192
[arXiv:hep-ph/9310247 [hep-ph]].



\bibitem{Locher:1993cc}
M.~P.~Locher, Y.~Lu and B.~S.~Zou,
Z. Phys. A \textbf{347} (1994), 281-284
doi:10.1007/BF01289796
[arXiv:nucl-th/9311021 [nucl-th]].



\bibitem{Li:1996yn}
X.~Q.~Li, D.~V.~Bugg and B.~S.~Zou,
Phys. Rev. D \textbf{55} (1997), 1421-1424
doi:10.1103/PhysRevD.55.1421

\end{thebibliography}
\end{document}